\documentclass[aps,prl,twocolumn,showpacs,preprintnumbers,amsmath,amssymb,superscriptaddress]{revtex4}
\usepackage{graphicx}
\usepackage{dcolumn}
\usepackage{bm}
\begin{document}

\title{Coleman meets Schwinger!}

\author{Lorenzo Cornalba}
\affiliation{Dip. di Fisica e Sez. INFN, Universit\'a di Roma ``Tor Vergata'',  Roma, Italy}
\author{Miguel S. Costa}
\affiliation{Centro de F\'\i sica do Porto e Departamento de F\'\i sica, Faculdade de Ci\^encias da Universidade do Porto, Portugal}
\author{Jo\~ao Penedones}
\affiliation{Centro de F\'\i sica do Porto e Departamento de F\'\i sica, Faculdade de Ci\^encias da Universidade do Porto, Portugal}

\begin{abstract}
It is well known that spherical D--branes are nucleated in the presence of an external RR electric field. 
Using the description of D--branes as solitons of the tachyon field
on non--BPS D--branes, we show that the brane nucleation process can be seen as the 
decay of the tachyon false vacuum. This process can describe the decay of flux--branes
in string theory or the decay of quintessence potentials arising in flux compactifications.
\end{abstract}
\preprint{CFP--05--01}
\preprint{ROM2F/2005/01}
\maketitle

One of the most beautiful results in Quantum Field Theory, derived by Schwinger \cite{Schwinger}, 
is that pairs of charged particles are produced in an external electric field. This mechanism
can be generalized to the nucleation of spherical branes in theories with $p$--form gauge 
potentials and was studied in \cite{Teitelboim} using semiclassical instanton methods,
starting from the Nambu--Goto brane action minimally coupled to the gauge field.
String theory has many extended charged objects associated to massless gauge fields and therefore
analogous computations of brane nucleation rates can be performed.

According to Sen \cite{Senrev}, D--branes can be thought of as tachyon solitons. It is therefore natural 
to ask if the brane nucleation process can be described in this language. Consider, in particular,
the tachyon kink solution on a non--BPS D$p$--brane that interpolates between the $T=\pm\infty$ vacua 
of the tachyon field and which describes a BPS D$(p-1)$--brane.
By turning on an  external RR electric field
$$
dC_p = E \,\,dt\wedge dx^1 \wedge \cdots \wedge dx^p
$$
parallel to the worldvolume of a non--BPS D$p$--brane, we shall see that the tachyon vacuum
degeneracy is lifted and that the brane nucleation corresponds to the decay of the tachyon false vacuum.
Hence Coleman's analysis  of the decay of the false vacuum \cite{Coleman} describes Schwinger's nucleation process.
Similar methods were used in \cite{Hashimoto} to describe brane/anti--brane decay by creation of a 
throat \cite{throat}.

We shall take as closed string background for our computations flat space with a small RR
electric field and we shall neglect the backreaction on the closed string geometry. The electric
field $E$ will in general distort the geometry on length scales larger than $1/E$, therefore
our analysis will be valid only within this scale. There are two simple settings to keep in mind.
Firstly, whenever the directions transverse to the electric field are not compact, the
background geometry is that of a flux--brane \cite{fluxbranes}. The electric field self--gravitates
on the length scale $1/E$ and the non--BPS brane is placed at the center of the flux--brane,
which is a gravitationally stable point. In this language the decay of the flux--brane 
towards flat space is explicitly described as the decay of the open string 
false vacuum of many non--BPS D-branes. Secondly,  whenever the transverse directions are
compact, the electric field will give rise to a quintessence potential in the 
compactified theory \cite{CCK}. The decay of the tachyon field will then describe 
the decay of the quintessence potential. This process generalizes the dynamical decay of the
cosmological constant of \cite{BrownTeit}. We shall neglect the effect of the expansion of the 
universe in the tachyon dynamics.

Throughout this paper we neglect closed string effects by taking $\alpha'E^2\ll 1$ and
$g_s\ll 1$. Therefore we can consider the open string dynamics independently, ignoring the
backreaction on the closed string fields. The effects in the closed string 
fields due to the vacuum decay can be computed to leading order and describe a change of order $g_s$.
We work in units such that $\alpha'=1$.

Recall the form of the classical action for the tachyon field on a non--BPS 
D$p$--brane \cite{Senrev}
\begin{eqnarray}
S &= &-\int d^{p+1}x\, V(T) \sqrt{1+\eta^{\mu\nu}\partial_\mu T\partial_\nu T}
\nonumber
\\
&&+\int W(T)\,dT\wedge C_p\ ,
\label{action}
\end{eqnarray}
where the massless fields on the brane are consistently set to zero and $\mu,\nu=0,1,\cdots,p$.
We consider flat  space--time with a constant dilaton, but we 
allow for the presence of a RR $p$--form potential. 
Several properties of the functions $V(T)$ and $W(T)$ are known.
Both are even functions and behave asymptotically as $e^{-|T|/\sqrt{2}}$.
Moreover, $V(0) = \tilde{T}_p$ is the tension of the non--BPS D$p$--brane,
which is related to the tension $T_p$ of a BPS D$p$--brane by $\tilde{T_p}=\sqrt{2}\,T_p$.
Finally, the fact that the tachyon kink solution 
represents a BPS D$(p-1)$--brane gives the additional requirements
\begin{equation*}
\int_{-\infty}^{\infty} dT\, V(T) = \int_{-\infty}^{\infty} dT\, W(T) =
T_{p-1}\ ,
\end{equation*}
so that the tension and charge of the soliton are correctly normalized.
The qualitative results in this note will not depend on the  specific form of the functions 
$V(T)$ and $W(T)$. However, for specific examples we shall use the explicit form \cite{Senrev}
\begin{equation}
V(T) = W(T) = \frac{\tilde{T}_p}{\cosh\left(T/\sqrt{2}\right)}\ .
\label{choice}
\end{equation}

To analyze the tachyon dynamics in the presence of an external RR electric field
it is convenient to integrate by parts the Wess--Zumino term in the action.
Defining 
\begin{equation*}
Z(T) = - \int_{T}^{\infty}dT'\, W(T')\ ,
\end{equation*}
and dropping the boundary term, one obtains simply
\begin{equation*}
S_{WZ} = -\int Z(T)\,dC_p\ .
\end{equation*}
The function $Z(T)$ was defined so that $Z(\infty) = 0$ and $Z(-\infty) = - T_{p-1}$. 
From the asymptotics of $W(T)$ it follows that $Z(T)$ approaches its
asymptotic values at $T=\pm\infty$ as $\mp\,e^{-|T|/\sqrt{2}}$.
The tachyon effective action in the presence of an electric field can then be written as
\begin{equation*}
S=-\int d^{p+1}x \left[V(T) \sqrt{1+\eta^{\mu\nu}\partial_\mu T\partial_\nu T}
+E\,Z(T)\right],
\end{equation*}
and the corresponding classical effective potential is
\begin{equation*}
U(T) = V(T) + E\,Z(T)\ .
\end{equation*}
The constant of integration in the function $Z(T)$ above was fixed so that $U(\infty) = 0$.
From the asymptotics of $V$ and $Z$ it is 
clear that, for small electric field, both $T=\pm\infty$ are still vacua of the theory.
\begin{figure}
\includegraphics[width=0.90\columnwidth,keepaspectratio,clip]{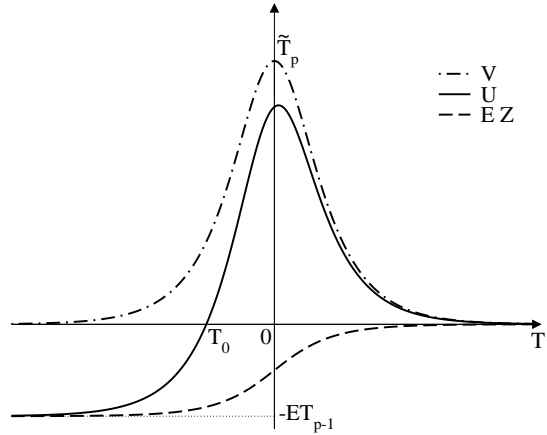}
\caption{\label{fig:potential} Tachyon potential $U(T)$ and functions $V(T)$ and $EZ(T)$.}
\end{figure}
However, since $U(-\infty) =  - E\, T_{p-1}$,
$T=+\infty$ is a metastable vacuum and will tunnel to the true vacuum at $T=-\infty$ (see FIG. 1).
On the other hand, from the asymptotics of $V$ and $Z$ it is also clear that, as one increases
$E$ above a critical value, $T=+\infty$ ceases to be a metastable vacuum. The
critical field depends on the specific form of $V$ and  $W$, but it will be 
of order the string scale. For the particular choice (\ref{choice}) the critical
value is $1/\sqrt{2}$.

In the above language it is easy to check that the kink representing a D$(p-1)$--brane
feels a force due to the external field. Consider a configuration depending on a single variable
$x^p$ with $T(\pm\infty)=\pm\infty$. Then, under a translation $x^p\to x^p+\delta x^p$, the 
change in the energy density is $\delta x^p E U(-\infty)$, which gives the constant pressure
$E T_{p-1}$.

Since  $T=+\infty$ is metastable one expects that 
there is a finite rate per unit time and volume for the nucleation of bubbles of the true vacuum 
inside the false vacuum. Furthermore, we see that the surface of the bubble is the locus where the 
tachyon field interpolates between the two minima, and should therefore be interpreted as a 
spherical D$(p-1)$--brane. We will estimate the nucleation rate using the standard semiclassical 
methods developed by Coleman. This amounts to finding the bounce: the non--trivial classical solution 
of the Euclidean equations of motion, which approaches the false vacuum at infinity and has minimal 
Euclidean action $S_{0}$. Then  $\Gamma/V \simeq \Delta\, e^{-S_{0}}$
is the nucleation rate per unit time and $p$--volume, with $\Delta$ a pre--factor. This pre--factor is related
to the determinant of the open string field fluctuations around the bounce. Since the tachyon classical
action is obtained by eliminating the massive open string fields via their equations of motion, the naive
pre-factor computed from the action (\ref{action}) would neglect fluctuations of the massive string modes.

Since the kinetic term in the action is not canonical, one should verify the validity of standard
instanton methods for the  Born-Infeld Euclidean action
\begin{equation*}
S_{E}=\int d^{p+1}x\, \left[V(T) \sqrt{1+(\partial T)^{2}}
+E\,Z(T)\right].
\end{equation*}
Let us call the bounce solution $\bar{T}$ and define a one--parameter family of 
functions $T_{\lambda}(x)\equiv\bar{T}(x/\lambda)$. One can easily show that 
\begin{eqnarray*}
\left.\frac{d}{d\lambda}\,S_{E}\left[T_{\lambda}\right]\right|_{\lambda=1} =&&
(p+1)\,S_{E}\left[\bar{T}\right] 
\\
&&
-\int d^{p+1}x\,\frac{V(\bar{T})(\partial \bar{T})^{2} }{\sqrt{1+(\partial \bar{T})^{2}}} \ .
\end{eqnarray*}
Since $\lambda=1$ is a stationary point of the action it follows that
$S_{E}\left[\bar{T}\right]>0$. One may also compute the second derivative
at $\lambda=1$ and easily show that it is negative.
This fact guarantees the existence of a negative eigenmode in the quadratic
fluctuations around the bounce. One also expects that the bounce is spherically symmetric
and that it has precisely one negative eigenmode.

To simplify the analysis of the dynamics of the bounce, we shall work with
potential functions ${\cal V},{\cal U},{\cal Z}$, which are obtained from the original 
$V,U,Z$ by rescaling them by $2/T_{p-1}$.
Start by defining a new field 
\begin{equation*}
\Phi=\int_0^T\,{\cal V}(T')\,dT'\ ,
\end{equation*}
such that $T=\pm\infty$ corresponds to $\Phi = \pm 1$. For the particular choice
(\ref{choice}), one has
\begin{equation}
{\cal V}=\frac{\sqrt 2}{\pi}\cos\left(\frac{\pi}{2}\,\Phi\right)\ ,\ \ \ \ \ \ \ \ \ \
{\cal Z}=\Phi - 1\ .
\label{scaledchoice}
\end{equation}
For a Euclidean solution with maximal symmetry
$\Phi=\Phi(r)$ where $r^2=x_\mu x^\mu$, the effective one--dimensional action reads
\begin{equation*}
I_E=\int_0^\infty dr\,r^p
\left[\sqrt{{\cal V}^2+\dot{\Phi}^2} + E\,{\cal Z}\right]\ ,
\end{equation*}
where dot denotes the radial derivative.
The Euclidean action is $S_E=I_EA_pT_{p-1}/2$, with $A_p$ the volume of the $p$--sphere.

The equations of motion turn out to be simpler in the Hamiltonian formulation. The conjugate
momentum to the field $\Phi$ is related to
\begin{equation*}
\frac{1}{r^p}\,\frac{\partial L}{\partial \dot{\Phi}}
=\frac{\dot{\Phi} }{\sqrt{{\cal V}^2+\dot{\Phi}^2} } 
\equiv  \Pi\ .
\end{equation*}
Thus, since ${\cal V}$ is non--negative, the variable $\Pi$ is also bounded between $-1$ and $1$.
Finally, the equations of motion reduce to the following dynamical system
\begin{eqnarray}
\dot\Pi &=& {\cal V}'(\Phi) \sqrt{1-\Pi^2} + E\,{\cal Z}'(\Phi) - \frac{p}{r}\,\Pi\ ,
\nonumber
\\
\dot\Phi&=&\frac{{\cal V}(\Phi)\,\Pi}{\sqrt{1-\Pi^2}}\ .
\label{dynamics}
\end{eqnarray}
Although for $p>0$ the system is not conservative, it is still useful to consider the Hamiltonian
\begin{equation*}
-\frac{1}{r^p}\,H={\cal V}(\Phi)\sqrt{1-\Pi^2} + E\,{\cal Z}(\Phi) \equiv K\ .
\end{equation*}
Using  (\ref{dynamics}) it is easy to show that 
\begin{equation*}
\dot{K} = \frac{p}{r} \frac{{\cal V}(\Phi)\Pi^2}{\sqrt{1-\Pi^2}} \ge 0 \ .
\end{equation*}
Thus, it is instructive to consider the $K$ contour plot, since $K$ never decreases along 
the $(\Phi(r),\Pi(r))$ trajectories (see FIG. 2). Spherical symmetry demands $\Pi(0)=0$
and therefore all trajectories start from the $\Pi=0$ horizontal axis.
\begin{figure}
\includegraphics[width=0.90\columnwidth,keepaspectratio,clip]{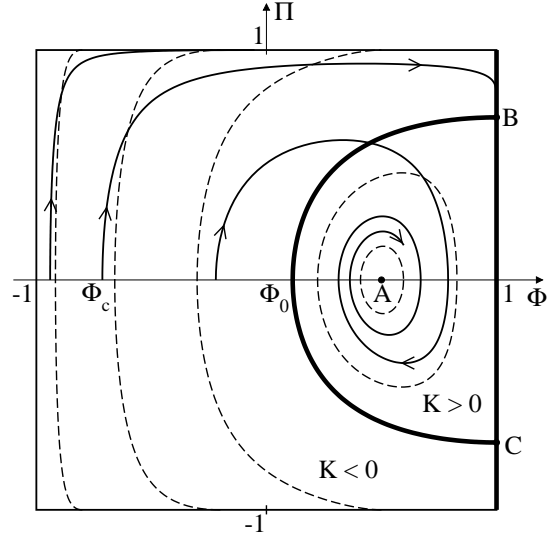}
\caption{\label{fig:contour} Contour plot of $K$ and typical trajectories for
$E>E_c$. The thick  line is the $K=0$ contour and the dashed lines are other $K$ contours.
The line starting at $\Phi_c$ is the bounce for $p>0$ and the other two lines
are typical trajectories. When $E\le E_c$ the bounce degenerates to the boundary of phase space.}
\end{figure}

Consider first the conservative case $p=0$, where the trajectories follow lines of constant $K$ clockwise.
The bounce solution is the trajectory with $K=0$, with the initial value $\Phi_0$ defined by the equation 
$K(\Phi_0,0)={\cal U}(\Phi_0)=0$. This trajectory approaches the vertical line $\Phi=1$ as $r\to\infty$ (point B in FIG. 2).
The corresponding Euclidean action is given by the integral
\begin{equation*}
S_0 = T_{p-1}\int_{\Phi_0}^1 d\Phi\sqrt{1-\left(\frac{E\,{\cal Z}}{{\cal V}}\right)^2}  \ .
\end{equation*}
In the limit $E\to0$ the above action remains finite and describes a $D/\bar{D}$--instanton pair. 
Trajectories that start with $\Phi(0)$ between $\Phi_0$ and $1$ are periodic and 
those that start  between $-1$ and $\Phi_0$ end in finite $r$ at the horizontal line  $\Pi=1$.

Next we consider the case $p>0$. Now all trajectories that enter the $K>0$ region will 
inevitably approach point $A$ in FIG. 2 at $r=\infty$.
This will happen for $\Phi(0)$ greater than some critical value $\Phi_c$ satisfying $-1\le\Phi_c<\Phi_0$.
On the other hand, trajectories with $\Phi(0)<\Phi_c$ reach $\Pi=1$ in finite $r$ and can not be continued further.
The bounce is precisely the solution with  $\Phi(0)=\Phi_c$.
To understand the behavior of $\Phi_c$ as a function of the electric field $E$, let us look at the extreme solution
with $\Phi(0)=-1$. This trajectory stays in the  vertical line $\Phi=-1$  until it hits $\Pi=1$ at a finite radius $\bar{r}$.
Expanding the solution around this point, one can show that, for
\begin{equation*}
 E>E_c = \sqrt{\frac{p}{8}}\,\frac{1}{{\cal Z}'(-1)} \ ,
\end{equation*}
we necessarily have $\dot{\Pi}(\bar{r})>0$ and therefore the trajectory ends.
We conclude that for $E>E_c$ the bounce starts with $\Phi_c>-1$ and approaches point $B$ in FIG. 2 at $r=\infty$.
This behavior is analogous to the case $p=0$, where the electric field is always above $E_c$.
Although this solution is interesting, it is not clear if it survives string corrections.
As $E$ decreases to $E_c$ the bounce trajectory degenerates to the boundary of the phase space. 
For $E\le E_c$ the solution starts at $\Phi_c=-1$ and hits $\Pi=1$ at a radius $R=p/E$, where $\dot\Pi$ vanishes. 
Then the trajectory jumps discontinuously  to $\Phi=\Pi=1$ and descends vertically to point $B$ in Fig. 2.
As a function of $r$, the tachyon field $\Phi$ jumps discontinuously from the true to the false vacuum at
$r=R$, while $\Pi(r)$ is smooth. We conclude that in this case 
{\em the thin wall approximation is exact}. This is a consequence of the kinetic 
term of Born--Infeld type. The Euclidean action of the 
bounce can then be explicitly evaluated
\begin{equation*}
S_{0}=  \frac{A_{p}}{p+1} \, T_{p-1} \, R^{p}  \ .
\end{equation*}
This result was obtained by Teitelboim almost 20 years ago using first quantized $p$-brane 
mechanics \cite{Teitelboim}. Here we have 
derived it as the bubble decay of the tachyon false vacuum for the case of D-brane nucleation in string theory.

The time evolution of the nucleated branes is determined by the analytic continuation of the bounce solution. 
For $p>0$ a $D(p-1)$--brane appears as a $(p-1)$--sphere of radius $R$ and then expands with constant radial 
acceleration $1/R$ describing a hyperboloid in Minkowski space. In this case all the energy arising from the decay of the
false vacuum is transfered to the brane.
For $p=0$ there is no brane to absorb the energy
and therefore the tachyon field can not appear exactly in the true vacuum. 
In this case the tachyon field rolls down classically
from $T_0$ (see FIG. 1) to the true vacuum.

After nucleation the brane will act as a source for closed strings.
The tadpoles for the closed string fields can be computed by coupling the action (\ref{action}) to these fields.
Since the transverse coordinates of the original non--BPS brane vanish at the bounce, 
the only non--vanishing components of the sources for the metric and RR $p$--form field 
lie along the non--BPS brane world--volume.
A simple computation shows that the metric, RR $p$--form and dilaton tadpoles are  
\begin{eqnarray*}
T_{\mu\nu} &=&T_{p-1} \left( \frac{x_{\mu}x_{\nu}}{R^2} - \eta_{\mu\nu}\right)\delta(r-R)\,\delta_\perp (x)\ ,
\\
J_{\mu_{1} \cdots \mu_{p}}& =& T_{p-1}\,\frac{x^{\nu}}{R}\,\epsilon_{\nu\mu_{1} \cdots \mu_{p}}\,\delta(r-R)\,\delta_\perp (x)\ ,
\\
Q &=& T_{p-1}\,\delta(r-R)\,\delta_\perp (x)\ ,
\end{eqnarray*}
where $r^2 = x_{\mu}x^{\mu}$ is the analytic continuation of the 
Euclidean radial coordinate and $\delta_\perp (x)$ is a delta function 
on the transverse space. These are the sources that describe the decay of a RR flux $p$--brane. 
The linearized solution for the closed fields can be solved in terms of retarded Green functions and describes the 
radiation emitted by the expanding nucleated brane. This process can also be described in the framework of 
Euclidean quantum gravity using generalizations of the Ernst metric \cite{Dowker,CostaGutperle}. 
Here, instead of an effective closed string description, we used open strings to
describe the decay process and the sources for the emitted radiation.

When the transverse directions to the non--BPS brane are compact, the delta functions $\delta_\perp (x)$
in the above sources should be replaced by the inverse of the volume of the transverse space. The sources
induce a jump in the derivative of the dilaton, gauge potential and  metric. Moreover, the dilaton
will radiate. In the dimensionally reduced theory the jump
in the electric field corresponds to the decay of a quintessence potential. It would be interesting to
consider the explicit form of the background geometry and analyze the decay process throughout the
cosmological evolution.
 
The techniques explored in this paper can also be applied to black hole physics. In particular, one
could try to make predictions regarding the emission of branes by non--extremal black holes in string
theory. 

\bigskip
This work was supported in part by INFN, by the MIUR--COFIN contract
2003--023852, by the EU contracts MRTN--CT--2004--503369,
MRTN--CT--2004--512194 and MERG--CT--2004--511309, by the INTAS contract 03--51--6346, by the NATO
grant PST.CLG.978785 and by the FCT contracts POCTI/FNU/38004/2001 and  POCTI/FP/FNU/50161/2003.
L.C. is supported by the MIUR contract ``Rientro dei cervelli'' part VII and J.P.
by the FCT fellowship SFRH/BD/9248/2002.

\end{document}